# Software for producing trichromatic images in astronomy

Sébastien Morel[1,2] and Emmanuel Davoust[1]

[1] : *UA 285, Observatoire Midi-Pyrénées, 14 Avenue Belin, F-31400 Toulouse, France*

[2] : *Ecole Supérieure d'Electricité, 2 rue Belin, F-57078 Metz Cedex 3, France*

**Abstract.** We present a software package for combining three monochromatic images of an astronomical object into a trichromatic color image. We first discuss the meaning of "true" colors in astronomical images. We then describe the different steps of our method, choosing the relevant dynamic intensity range in each filter, inventorying the different colors, optimizing the color map, modifying the balance of colors, and enhancing contrasts at low intensity levels. While the first steps are automatic, the last two are interactive.

## 1. Introduction

False-color images of astronomical objects, where different colors represent only different intensity levels in the image, are extremely common in the literature. They allow a much larger dynamic range in intensity to be displayed than, for example, grey-scale images. These images, of course, do not tell us anything about the colors of the objects.

On the contrary, "true" color images of astronomical objects, as they would be seen through the telescope, were the eye sensitive enough, are still quite rare. Astronomers have shown very little interest in producing such images; they prefer quantitative measures, such as $(B - V)$ color index maps, for example. The main reason probably is that the true colors of a planetary nebula or of a galaxy do not really mean anything, as they cannot be related to anything comparable on Earth. Another reason is that large telescopes no longer have eye-pieces through which one can gaze at the sky; they have conveniently been replaced by highly sensitive video or CCD cameras that allow astronomers to view the sky from their armchair, but in black-and-white, and they no longer know what the colors are.

One major exception is the solar system, as colors of planets can often be discerned at the telescope; these colors can also be related to the colors of the ground on Earth. The *Voyager* images of the solar system are in colors that match what we would see from close up. Another remarkable exception is Jim Wray's *Color Atlas of galaxies* (1988), which provides color images of over 900 galaxies, based on photographs taken in the (U, B, V) system. This atlas could very well be a reference for what galaxies should look like in colors, if the advent of CCD cameras had not shifted the standard filters toward the red, where these receivers are most sensitive. Finally, we should mention the growing interest of amateur astronomers; popular astronomy journals often display their successful attempts at taking color photographs of heavenly bodies.

Since color images of the sky by professional astronomers are now beginning to appear in the media, for example HST's first "corrected" view of M 100, computer programs for producing these color images must exist. But, surprisingly, very little information on the subject has appeared in the astronomical literature. The present paper is meant to fill this gap, and provides a general view of the subject, together with a detailed description of our method. We recommend that experts in image treatment go directly to Sections 6, 7, and 9, which contain the most original work.

## 2. The meaning of "true" colors

There are several methods for producing color images. In one method, HSI, the color of each image pixel is defined by its Hue (the color's peak wavelength), Saturation (the hue's purity) and Intensity. In another, RGB, the image is decomposed into three primary colors, Red, Green and Blue. The colors perceived by the human eye can largely be reproduced by the combination of three primary colors (for a non-technical presentation of color perception of the eye, see Robertson, 1992), and, for this reason, this method is used in computer and television color screens. This is also the method adopted here.

A first problem, when aiming at obtaining "true" color images by this method, is that the combination of filter transmission and camera spectral response used in astronomy does not necessarily match those used, say, by camcorders to successfully reproduce scenes in colors. One could, of course, imagine some kind of color equations to go from one system of filters to the other, but these equations would obviously depend on the nature of the object.

Another problem arises for emission-line objects. Because there is overlap in the domains of sensitivity of the three kinds of cones in the human eye, it is capable of perceiving the color of monochromatic objects to a certain extent. This is not the case for the set of wide-band filters commonly used in astronomy; one has to use images taken with narrow-band filters to reproduce the colors of monochromatic objects properly. This is why diffuse emission nebulae and planetary nebulae are tricky objects to reproduce in colors. Any amateur (or old professional) astronomer will tell you that the Orion nebula is greenish; but how many times have you seen it displayed in different shades of pink or red, simply because the [OIII] emission at 5007Å falls in between two wide-band filters!

In short, it is probably illusory to try and reproduce the true colors of astronomical objects for the time being.

Incidentally, in this paper a monochromatic image is an image obtained using a filter of any width, as opposed to a trichromatic image, while a monochromatic object is one that mainly emits one or several emission lines.

The next best thing that one can do is to produce the colors of the objects as they would be seen if our eyes had the same spectral response as the combination of camera and filter set used. This amounts to calibrating the relative intensities in the three monochromatic images, for example by measuring the response of the receiver to a source of given uniform (white) intensity seen through each of the three filters, and then correcting two of the three monochromatic images accordingly.

But this method, which is still realistic, will lead to disappointing results, because the colors of astronomical objects are mostly unsaturated, and their color images tend to be very pale; this is also true for many emission-line objects, as they tend to be immersed in white stellar light. For the whole truth about stellar colors, we refer to Steffey (1992). This is what we experienced with galaxies; calibrating monochromatic images with published aperture photometry of the object does not give aesthetically pleasing results. Furthermore, galaxies have such a large dynamic range that the outer parts of disks of spiral galaxies are barely perceptible, or else the central regions are completely overexposed and white.

These shortcomings led us to abandon any pretense at producing colors that are correct in some sense, in favor of finding by trial and error the image that looked the most pleasing, or the closest to what one expects the object to look like. At the same time, we had to drop the label of true color image.

## 3. Limitations of computer screens for displaying colors

When designing a method for producing color images, one first has to take into account the resources of computer screens for displaying them. Video boards generally use the pixmap format and its associated color map for displaying color images on computer screens.

In the pixmap format, and unlike bitmap, each pixel is no longer represented by the three intensities of the three color components that produce the true color, but by an integer which points to the corresponding color in the color map. In our case, the pixmap values range from 0 to 255, because most computers only allow a limited range of colors, generally 256, to be displayed simultaneously on the screen. This is true for PC-compatibles as well as for Unix workstations. Special video boards do allow wider ranges of colors; but, since they are not (yet) widespread, there is at present little point in producing images for such systems.

The pixmap is linked to a color map, which associates a triplet of RGB color intensities with each value of the pixmap. This color map is coded with 6, 8, or 16 bits per color component. This means that the 256 colors of the color map can be chosen from a very wide selection of colors. On PC-compatibles, for example, each of the three color components is coded on 6 bits, and there are consequently $64^3$ possible colors. On Unix workstations, the X11 system handles colors coded on 16 bits, and the choice is thus considerably larger, $65536^3$.

One way of circumventing the limitation of 256 colors is to code each image pixel with four pixels on the screen, one for each of the three primary colors and one in grey, whose intensity is a combination of the RGB intensities of the pixel. There is no need for compression of color dynamics, since the eye makes the synthesis of the colors displayed on the screen; the range of colors thus displayed is $64^3$.

There are several problems with this method. The first one is that color images are four times larger than the original monochromatic images, and it is the spatial, not the color resolution, of standard computer screens that becomes a limitation. Another problem is that, when one primary color dominates on an image pixel, 2 or 3 of the 4 corresponding screen pixels will be dark, and the eye is likely to catch the pattern that appears on the screen in the form of thin dark lines. Such patterns are also likely to produce Moiré effects when the image is printed. Finally, such color images cannot be stored directly in TIFF or any other standard output format; one could consider making a postscript file, which is of limited use.

We thus adopt the pixmap format, and do not consider further the alternative of 4 screen pixels per image pixel.

## 4. A new software package for producing trichromatic astronomical images

We had essentially two motivations for creating new software for producing color images. The first was to automatize most of the process, in particular the choice of the dynamic range. Expertise is generally required for selecting the relevant range in the three colors, without which one spends a lot of time before reaching a satisfactory image. The second motivation was to adapt the method to the specific nature of astronomical images. Galaxies, for example, have a large dynamic range, and we wanted to be able to enhance the low intensity levels, in order to see the whole galaxy in colors at once. Other advantages of such a dedicated software package include the choice of convenient image input and output formats, and the possibility of modifying the code to improve or add certain tasks.

We now describe the computer code in detail.

## 5. Choice of the relevant dynamic range

Each of the three monochromatic images spans a certain range of intensities. The lowest intensities correspond to the sky level, and the brightest ones to the nucleus of an extended object, or to some bright stars in the field, and there is no relevant information outside this range of intensities. It is thus important to rescale the intensities in the three images in order to reduce their dynamics.

The first step is to plot a histogram (Fig. 1) of the number of image pixels with a given intensity in each monochromatic image. The distribution is skewed; the sky background (shaded area) is in the low-intensity part of the histogram, and the relevant information is in the long tail toward high intensities.

Two limits, $I_{min}$ and $I_{max}$, have to be adopted. The first one should be as close as possible to the mean value of the intensity of the sky background. We found that the median of the distribution is a good choice. $I_{min}$ is thus such that half the pixels of the image have an intensity less than $I_{min}$.

The choice for $I_{max}$ is more difficult, and depends on the nature of the astronomical object that has been imaged. Choosing the brightest intensity in the image is generally not a good choice, because this intensity could correspond to a cosmic ray impact, or to a field star, or be extreme in some other way. The best way is to compute a sliding mean frequency at high intensities, and to choose $I_{max}$ such that there is an average of n pixels per intensity unit around $I_{max}$, where n is between 3 and 10 pixels for CCD images of standard size. It is, in fact, better to choose the lower value, as one can always reduce the dynamics further, but not increase it, in the next steps.

Values of 50% of sky background and of 3 pixels per intensity unit have been adopted as default values for parametrizing the two intensity limits. If the final image is not satisfactory (e.g. sky background and/or center of the object too bright or too faint), the program can be run again with explicitly given values, for example 55% and 6.

We thus transform the intensities $I$ in each monochromatic image into normalized intensities $i$ in the following way :

$$i = I_{min} \quad \text{for} \quad I < I_{min} \tag{1}$$

$$i = I_{max} \quad \text{for} \quad I > I_{max} \tag{2}$$

and

$$i = 127 \times \frac{I - I_{min}}{I_{max} - I_{min}} \quad \text{in} \quad \text{between} \tag{3}$$

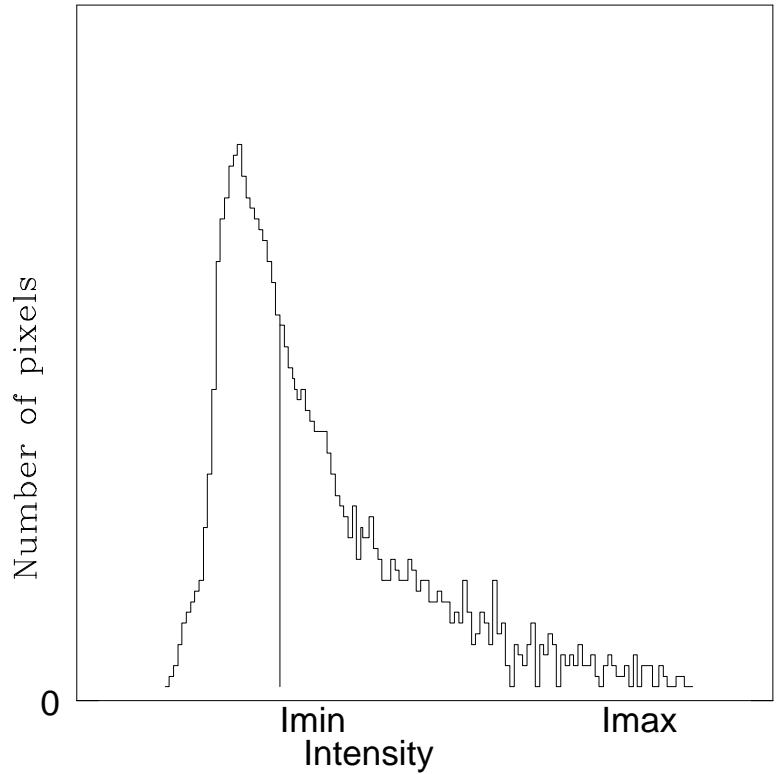

Figure 1: *histogram of pixel intensities*

This normalization produces monochromatic images that have very similar intensities. It does not take into account the true color of the object, and the resulting trichromatic image will in principle be rather white or greyish. The reason for this choice is to retain the largest possible meaningful dynamics, which can be further reduced at will later. The $I_{max}$ of each color component will again be modified after color optimization.

The resulting image is temporarily stored in bitmap format, where each pixel is coded on 3 bytes (each color component has a value between 0 and 127), which provides a fair colorimetric resolution, but without an excessive number of color classes. While the total number of possible colors allowed with this coding is $128^3$, that is over 2 million, a typical number of colors encountered in our astronomical images is 20 000. This number of 20 000 colors is repeatedly quoted in the remainder of the paper, but it is understood that it could be anywhere between $10^4$ and $10^5$, depending on the images. By color we mean a given triplet of RGB colors.

## 6. Inventory of the different colors and their weight

In order to proceed, we need to make an inventory of the different colors that have been produced, and to determine the weight of each color (i.e. the number of pixels that belong to it). Sorting between $3 \times 10^5$ and $10^6$ pixels into about 20000 colors is not trivial, and the difficulty is to make the procedure as efficient as possible.

In practice, this inventory is made at the same time as the temporary bitmap is created. One scans the image, pixel per pixel. For each pixel, one scans the list of colors. If the triplet already exists, its weight is increased by one; if it does not exist, it is created and given a weight of one.

If there is only one list of colors, the number of colors to be scanned increases roughly in linear progression as one proceeds from one row to the next, thus approximately like $N_{row}(N_{row}+1)/2$, where $N_{row}$ is the number of rows in the image.

The trick is to divide the list into $N$ lists evenly covering the color range, so that all lists are of comparable sizes, and that one color only belongs to one list. Each list then has $128^3/N$ possible colors, and the total scanning time is reduced by a factor $N$. This reduction outweighs the added overhead of the 5 multiplications, 2 additions and 3 mod functions necessary for computing the index $k$. A list is defined by its index $k$ as a function of the three values $i_R$, $i_G$ and $i_B$ of the three color components :

$$k = L^2(i_R \, mod \, L) + L(i_G \, mod \, L) + i_B \, mod \, L \qquad (4)$$

where $L^3 = N$, and $mod$ stands for the modulo operator; the index $k$ ranges from 0 to $N - 1$.

This way of selecting $N$ lists prevents one or several lists from having many more elements than the others, which would otherwise occur, for example, if the image is predominantly red. At the end of the procedure, the $N$ lists are simply merged into one.

In practice we found that $N = 512$, and thus $L = 8$, was a good choice.

## 7. Optimizing the choice of colors

The next task of the computer code is to gather the 20000 colors that have been inventoried into 256 color classes, in order to bring the number of triplets to 256. Afterwards, it becomes trivial to associate one color of the color map to each triplet, and thus to each pixel. At the end of this optimization, the image is stored in pixmap format.

We first deal with the color black; it is assigned to the sky background, and we are thus left with 255 colors.

Selecting the most frequent triplets to define the different classes is not a good method, because one may then end up with several colors that are close to each other, perhaps so close that the eye cannot tell the difference, and one may ignore other colors, which are infrequent, but which would provide a valuable nuance in colors. On the contrary, we try to group colors that are not very different into the same class.

Among the methods used for such a color quantization (which is a "vector quantization" issue), we have chosen a Linde-Buzo-Gray (LBG) algorithm (Linde et al. 1980), also known as the K-means algorithm. There are faster quantizers, like Bragg's (1992) reduction filter, but such a method should not be used in our context, because it generates noise which might drown the texture of objects like galaxies. Furthermore, because our software is not a real-time application, the speed of execution is less important than the quality of the result.

The main problem with any LBG quantizer lies in the choice of the initial codebook (here, the color map). A random choice does not lead to an optimal color map after running the LBG algorithm. Heckbert (1982) proposed a method which he calls "median-cut algorithm", and which has the advantage of creating classes of colors with equal numbers of pixels. However, according to Orchard and Bouman (1991), this method is greatly influenced by the initial palettes, and leads to a local minimum of colorimetric bias (see Section 7.4) which is not necessarily a global minimum.

Feeling that the LBG algorithm is slow and does not improve markedly "good" initial color maps, Orchard and Bouman (1991) have proposed a quantizer which does not use LBG, which they call "binary tree algorithm". This algorithm has recently been improved by Balasubramanian et al. (1994).

Ranking the quality of the result before the speed of execution, we implemented an efficient and original initial-color-map designer, in three steps. We first gather all the triplets into a little over 255 classes by a crude method, then assign a barycenter and a weight to each class by LBG, and finally bring the number of classes to exactly 255 by the elegant, but more time consuming, method of hierarchical clustering. We feel that this combination of several methods is a good compromise for our purpose.

*7.1 Defining the initial classes*

The LBG method requires that RGB space be divided into N inital classes, each with a barycenter and a weight. We obtain the N initial classes in the following way.

We place the colors in three-dimensional RGB space, each coordinate being the value of the corresponding component of the color triplet. Each point in this space is given a weight proportional to the number of image pixels belonging to that specific color. The rule of the game is to divide this space into 256 or more cubes, each of about the same total weight (each containing approximately the same number of image pixels).

The initial cube, which contains all the points, has a dimension of $128^3$ by construction. This cube is divided into 8 equal parts, each of which is assigned a total weight which is simply the sum of the individual weights.

According to its weight, each of the 8 smaller cubes is then either ignored, left as it is to represent a final class of colors, or divided again. In the first case, when the weight is below the survival threshold ($ST$), there will be no color class associated with that cube, and the pixels belonging to that cube will each be given the color of an adjoining pixel in real space (not in RGB space). In the second case, when the weight is between $ST$ and the fecundity threshold ($FT$), all pixels in this cube are assigned to that class. In the last case, when the weight is larger than $FT$, we do one more iteration : the cube is divided into 8 equal parts, the weight of each smaller cube is computed, and so on.

As initial values for $ST$ and $FT$, we take 0.01% of the total number of pixels, and 1.5 times the mean class weight. The latter is defined as the number of meaningful pixels (50% of the total, since the other half of the image is sky background and given the color black) divided by the number of colors others than black (i.e. 255).

$$ST = 0.01 \times N_{rows} \times N_{cols} \qquad (5)$$

$$FT = 1.5 \times \frac{0.50 \times N_{rows} \times N_{cols}}{255} \qquad (6)$$

where $N_{rows}$ and $N_{cols}$ are the numbers of rows and columns of the image.

If these initial values lead to less than 255 classes, a second cycle of iterations is performed, with $ST$ and $FT$ divided by 2.

Heckbert's (1982) median-cut algorithm is another way of obtaining the N initial classes of colors. We did not check whether or not it is better than ours.

## 7.2 Assigning each pixel to a class by LBG

We now have N classes. Each class $X_n$ initially has zero weight, and a barycenter defined by the geometrical center of the corresponding cube in RGB space. We proceed to assign each triplet $x_i$ with weight $p_i$ (equal to the number of image pixels having that color) to a class $X_n$ if the distance of that color to the barycenter of $X_n$ is smaller than its distance to any other class $X_m$.

$$d(x_i, X_n) < d(x_i, X_m) \tag{7}$$

where

$$d_{i,n} = \sqrt{(r_i - r_n)^2 + (g_i - g_n)^2 + (b_i - b_n)^2} \tag{8}$$

defines the distance between a point of index $i$ and one of index $n$.

The new class $X_n$ will have a new barycenter, which is the barycenter of $X_n$ and $x_i$, and a new weight, which is $p_n + p_i$. And so on.

In the end, the number of classes has not changed, but the barycenter (thus the color) and weight of each of them has, for the better.

## 7.3 The final selection of classes by hierarchical clustering

The above algorithm provides us with N color classes, where N is between 256 and about 300. These have to be reduced further, by the method of hierarchical clustering. We still use RGB space, and define the distance $d_{i,j}$ between classes $X_i$ and $X_j$ as above. We then compute the N(N-1)/2 distances between all N color classes and put them in a N×N matrix.

We next find the smallest distance, say $d_{m,n}$, and merge classes $X_m$ and $X_n$ into one class, which is assigned the color (i.e. the RGB coordinates) of the barycenter of the two. The matrix is thus reduced by one line and one column. We compute the new distances between the $N-1$ classes, and proceed until the size of the matrix is 255×255.

The time requested for computing all these distances is proportional to

$$\sum_{n=256}^{n=N} n(n-1)/2 = \frac{N^3 - N}{6} - 2763520 \tag{9}$$

This is one of the reasons why this method cannot be used right at the outset, with a large number of classes. Another reason is that the size of the N×N matrix becomes too large for the RAM of the computer.

## 7.4 A test of the color selection

The colorimetric bias $CB$ is a criterion for testing the quality of the color selection; it is simply the average displacement of individual pixels in RGB space, before and after optimizing the choice of colors to be displayed, and is defined as :

$$CB = \frac{1}{N_{rows} N_{cols}} \sum_{i=1}^{N} \sum_{j=1}^{M} \sqrt{(r_{ij} - r'_{ij})^2 + (g_{ij} - g'_{ij})^2 + (b_{ij} - b'_{ij})^2} \qquad (10)$$

where (r, g, b) and (r', g', b') are the coordinates in RGB space of each pixel before and after optimizing.

With the present algorithm, the $CB$ is typically of the order of 0.5 to 2.0 in a cube, 128 to a side. Such small nuances are barely perceptible to the human eye, and do not affect the fidelity of the image colors. This justifies adopting the present optimization method.

In practice, we have stored the bitmap after normalization, and the pixmap after optimization, and the $CB$ is computed from both maps. The $CB$ is not used in the later stages of the procedure; it is only useful for testing our algorithm and comparing it to others.

The different steps described so far for obtaining the color image are automatic; they do not require any expertise from the user. The next two steps, modifying the balance of colors, enhancing the contrast at low luminosity levels, are interactive.

## 8. Modifying the balance of colors

Because we have not calibrated the relative intensities between the three monochromatic images before normalization, the resulting color image tends to be whitish. One thus needs an *a priori* knowledge of the colors of the object in order to modify the balance of colors in such a way as to arrive at a "realistic" image. This has to be done interactively, by clicking on appropriate widgets. One first selects the component(s) to be modified, then the sense and the amplitude of the modification (± a factor 1.05 or 1.20). These two factors were adopted in order to allow one to display a wide range of colors in rapid succession.

A second modification is sometimes deemed necessary, namely that of the $I_{min}$ of one component, if a strong amplification of contrasts at low-luminosity levels shows the sky background to be slightly bluish or greenish for example. Again, this is done interactively, by clicking on a widget.

These two modifications reduce slightly the dynamics of the colors, and one could consider going back to optimizing colors with these new limiting intensities; we felt that such a complication of the computer code was not warranted.

## 9. Enhancing contrasts at low intensities

Certain astronomical objects, such as galaxies, present a very large dynamic range, which the unaided eye cannot completely perceive. A trick has to be used in order to reduce this range to a perceptible one. In false-color astronomical imaging, one often uses a logarithmic transformation; but it has the disadvantage of reducing contrasts at high intensity levels.

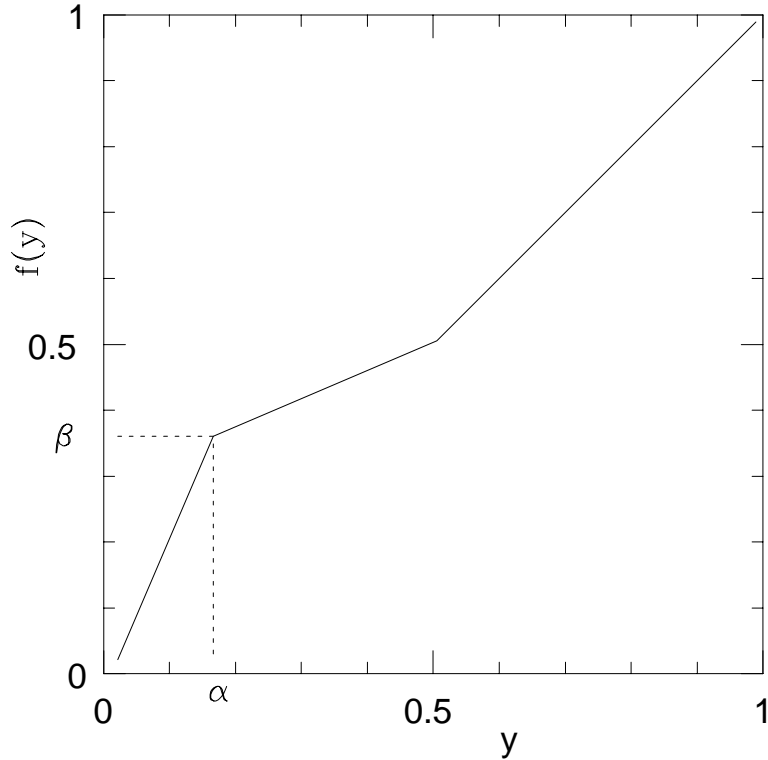

Figure 2: *transfer function for the luminance which enhances contrasts at low luminances*

We want a transformation which enhances low-luminosity levels without affecting bright ones. Because this transformation is non-linear, it cannot be applied indiscriminately to low-luminosity levels, without taking into account the intensities of the other components of a triplet. Finally, because it cannot be made automatically, this transformation should also be user friendly, so that the desired result be rapidly obtained.

We first define the luminance $y$ as the brightest of the three intensities in a color triplet. On Unix workstations, $y$ is coded from 0 to 65535; on PC compatibles with VGA board, from 0 to 255; the maximum of $y$ is denoted $y_{max}$.

We then define a transfer function $f(y)$ for the luminance as follows (see Fig. 2).

$$1. \quad y > y_{max}/2; \quad f(y) = y \qquad (11)$$

$$2. \quad \alpha < y < y_{max}/2; \quad f(y) = \frac{y_{max}/2 - \beta}{y_{max}/2 - \alpha} y + \frac{\beta - \alpha}{y_{max}/2 - \alpha} \times y_{max}/2 \qquad (12)$$

$$3. \quad y < \alpha; \quad f(y) = \frac{\beta}{\alpha} y \qquad (13)$$

The derivative of $f(y)$ represents the contrast. In the first case, the slope of the transformation is unity, and the contrast is unchanged. In the second one, the slope is smaller than unity, and the contrast is reduced. In the last one, the slope is larger than unity, and the contrast is enhanced. Thus, in order to gain contrast in one region, one necessarily loses contrast in another; we have chosen to leave the brightest region unaltered.

Initially, $\alpha = \beta = y_{max}/4$. The computer code allows us to modify $\alpha$ and $\beta$ interactively, the result of the transformation is directly applied to the the color map, and the image with the new colors is thus immediately visible on the screen.

If low intensity levels are still not sufficiently enhanced, one has to start the program anew, with a lower value of the sky background.

**10. Practical information on the software package**

The computer code was written in C langage on an IBM RISC 6000 workstation with the AIX operating system; it uses X Window (Jones, 1989), and thus needs Xlib; the source has about 2000 lines. It can be obtained by e-mail from davoust@obs-mip.fr. There is no help on-line and we have not tested its portability.

One runs the program by typing a command line giving the generic name of the three input image files; one can also specify the percentage of background (50% by default) in the image and the required mean number of pixels (3 by default) at intensity $I_{max}$. It takes less than one minute to execute the first automatic steps (choice of dynamic range, initial classes, LBG, hierarchical clustering) for images of up to one million pixels. The image is then displayed and the user can modify the balance of colors, enhance low luminosity levels if necessary, and finally save the color image.

As input image format, we adopted the standard format for astronomical image files, which is the FITS format (for Flexible Image Transport System). Most telescopes are equipped with CCD receivers that produce images where the intensities are coded as integers (i.e. 16 bits per pixel). However, the FITS images produced after treatment by certain software packages may have 32-bit pixels. We thus allowed for input images with 8, 16 or 32 bits per pixel. FITS files have a header of 2880 records, and the image pixels, also of 2880 records, coded in bitmap. For more information on the FITS format, see Wells et al. (1981), and Grosbøl et al. (1988).

We adopted TIFF (for Tag Image File Format) as output format for our color images. A TIFF image has the following structure : a header of 8 bytes, a list of tags, tag data (optional), and the image in bitmap format; each pixel is coded on 24 bits, 8 per color. For a simple description of TIFF and a code in C-langage for reading and writing TIFF images, see Le Veilieppe (1992). Because TIFF is coded on 8 bits per color, and the color maps used under X11 are coded with 16 bits per component, one has to divide the values of the color map by 256 when converting to TIFF. Storing in TIFF format color images obtained on a workstation thus results in a minor loss of accuracy.

**Acknowledgements.** We thank David Okerson for help at the initial stage of this project, and an anonymous referee for very helpful comments on the manuscript.

# References

Balasubramanian, R., Allebach, J.P., Bouman, C.A., 1994, *J. Opt. Soc. Am.* A11, 2777

Bragg, D., 1992, *Graphic Gems Series*, Ed. Academic Press, 3, 20

Grosbøl, P., Harten, R.H., Greisen, E.W., Wells, D.C., 1988, A&AS 73, 359

Heckbert, P., 1982, *Computer Graphics*, 16, 297

Jones, O., 1989, *Introduction to the X Window system*, Ed. Prentice Hall

Le Veilieppe, P., 1992, *La Revue de l'Utilisateur PC*, n°78, 49

Linde, Y., Buzo, A., Gray, R.M., 1980, *IEEE Trans. Commun.*, COM-28, 84

Orchard, M.T., Bouman, C.A., 1991, *IEEE Trans. Signal Process.*, 39, 2677

Robertson, A., 1992, *Physics Today*, 45, n°12, p. 24

Steffey, P.C., 1992, *Sky & Telescope*, 84, 266

Wells, D.C., Greisen, E.W., Harten, R.H., 1981, A&AS 44, 363

Wray, J.D., 1988, *Color atlas of galaxies*, Cambridge U. Press.